\documentclass[12pt,english,floatfix,superscriptaddress,aps,prd,preprint,nofootinbib]{revtex4}
\usepackage{amsmath}
\usepackage{amssymb}
\usepackage{amsbsy}
\usepackage[a4paper, margin=1.8cm]{geometry}
\usepackage{amsfonts}
\usepackage{amsopn}
\usepackage{amstext}
\usepackage{graphicx}
\usepackage{amssymb}
\usepackage{amsfonts}
\usepackage{amsmath}
\usepackage{braket}
\usepackage{graphicx}
\usepackage[english]{babel}
\usepackage{color}
\usepackage{xcolor}
\usepackage{slashed}
\usepackage{esint}
\usepackage[dvips]{epsfig}
\usepackage[dvips]{graphicx}
\usepackage{float}
\usepackage{units}
\usepackage{textcomp}
\usepackage{placeins}
\usepackage{hyperref}             % Enable clickable links
\hypersetup{
    colorlinks=true,              % Colored links instead of boxed
    breaklinks=true,              % Allow links to break across lines
    citecolor=blue,               % Color for citation links
    linkcolor=[rgb]{0,0.5,0.9},   % Color for internal links
    urlcolor=red,                 % Color for URLs
    filecolor=green               % Color for file links
}

\usepackage{hyperref}
\usepackage{slashed}

\newcommand{\ie}{\begin{equation}}
\newcommand{\fe}{\end{equation}}
 \newcommand{\bq}{\begin{equation}}
 \newcommand{\eq}{\end{equation}}
 \newcommand{\bqn}{\begin{eqnarray}}
 \newcommand{\eqn}{\end{eqnarray}}

\begin{document}
%%%%%%%%%%%%%%%%%%%%%%%%%%%%%%%%%%%%%%%%%%%%%%%%%%%%%%%%%%%%%%%%%%%%%%%%

\title{Casimir phenomena in bumblebee gravity}

%%%%%%%%%%%%%%%%%%%%%%%%%%%%%%%%%%%%%%%%%%%%%%%%%%%}

%%%%%%%%%%%%%%%%%%%%%%%%%%%%%%%%%%%%%%%%%%%%%%%%%%%%%%%%%%%%%%%%%%%%%%%%%%%%%%%%%%%%%%%%%%%%%%%%%%%%%%%%%%%%%%%%%%%%%%%%%%%%%%%%%%%%%%%%%%%%

\author{D. S. Cabral}
\email{danielcabral@fisica.ufmt.br}
\affiliation{Programa de Pós-graduação em Física, Instituto de Física, Universidade Federal de Mato Grosso, Cuiabá, Brasil}

%%%%%%%%%%%%%%%%%%%%%%%%%%%%%%%%%%%%%%%%%%%%%%%%%%%%%%%%%%%%%%%%%%%%%%%%%%%%%%%%%%%%%%%%%%%%%%%%%%%%%%%%%%%%%%%%%%%%%%%%%%%%%%%%%%%%%%%%%%%%

\author{A. A. Ara\'{u}jo Filho}
\email{dilto@fisica.ufc.br}
\affiliation{Departamento de Física, Universidade Federal da Paraíba, Caixa Postal 5008, 58051--970, João Pessoa, Paraíba,  Brazil.}
\affiliation{Departamento de Física, Universidade Federal de Campina Grande Caixa Postal 10071, 58429-900 Campina Grande, Paraíba, Brazil.}
\affiliation{Center for Theoretical Physics, Khazar University, 41 Mehseti Street, Baku, AZ-1096, Azerbaijan.}

%%%%%%%%%%%%%%%%%%%%%%%%%%%%%%%%%%%%%%%%%%%%%%%%%%%%%%%%%%%%%%%%%%%%%%%%%%%%%%%%%%%%%%%%%%%%%%%%%%%%%%%%%%%%%%%%%%%%%%%%%%%%%%%%%%%%%%%%%%%%

\author{A. F. Santos}
\email{alesandroferreira@fisica.ufmt.br}
\affiliation{Programa de Pós-graduação em Física, Instituto de Física, Universidade Federal de Mato Grosso, Cuiabá, Brasil}

%%%%%%%%%%%%%%%%%%%%%%%%%%%%%%%%%%%%%%%%%%%%%%%%%%%%%%%%%%%%%%%%%%%%%%%%%%%%%%%%%%%%%%%%%%%%%%%%%%%%%%%%%%%%%%%%%%%%%%%%%%%%%%%%%%%%%%%%%%%%%%%%%%%%%%%%%%%%%%%%%%%%%%%%%%%%%%%%%%%%%%%%%%%%%%%%%%%%%%%%%%%%%%%%%%%%%%%%%%%%%%%%%%%%%%%%%%%%%%%%%%%%%%%%%%%%%%%%%%%%%%%%%%%%%%%%%%%%%%%%%%%%

\begin{abstract}

In this work, we analyze the Casimir effect associated with a massive, non–minimally coupled scalar field in static, spherically symmetric black hole spacetimes arising in bumblebee gravity. Three distinct solutions are considered, corresponding to different vacuum expectation value configurations of the Lorentz--violating vector field, including metric and \textit{metric--affine} scenarios. Finite--size effects are implemented through the Thermo Field Dynamics formalism by compactifying the radial direction, allowing the construction of renormalized vacuum expectation values of the energy–momentum tensor. Closed–form expressions for the Casimir energy and pressure are obtained in the massless limit as functions of the radial position of a spherical capacitor and the plate separation. Both observables depend explicitly on the bumblebee parameters and on the location of the apparatus relative to the horizon $R_0=2M$. In the weak--field regime, $r \gg R_0$, the standard flat–space behavior $E \propto -1/d^4$ is recovered. As $r \to R_0$, the Casimir energy vanishes while the radial pressure diverges. Inside the black hole, the interaction may alternate between attractive and repulsive regimes depending on the plate separation and on the Lorentz--violating couplings. A domain–dependent hierarchy among the three configurations emerges, with \textit{metric--affine} effects amplifying the interior vacuum energy, while configurations with simultaneous temporal and radial deformations dominate in the exterior region. Although all geometries share the same asymptotic Schwarzschild structure, their quantitative deviations become increasingly pronounced as the Lorentz–violating parameters grow.
\end{abstract}

\maketitle

%%%%%%%%%%%%%%%%%%%%%%%%%%%%%%%%%%%%%%%%%%%%%%%%%%%%%%%%%%%%%%%%%%%%%%%%%%%%%%%%%%%%%%%%%%%%%%%%%%%%%%%%%%%%%%%%%%%%%%%%%%%%%%%%%%%%%%%%%%%%%%%%%%%%%%%%%%%%%%%%%%%%%%%%%%%%%%%%%%%%%%%%%%%%%%%%%%%%%%%%%%%%%%%%%%%%%%%%%%%%%%%%%%%%%%%%%%%%%%%%%%%%%%%%%%%%%%%%%%%%%%%%%%%%%%%%%%%%%%%%%%%%%%%%%%%%%%%%%%%%%%%%%%%%%%%%%%%%%%%%%%%%%%%%%%%%%%%%%%%%%%%%%%%%%%%%%%%%%%%%%%%%%%%%%%%%%%%%%%%%%%%%%%%%%%%%%%%%%%%%%%%%%%%%%%%%%%

\section{Introduction} \label{S1}

Relativistic field theories are typically formulated under the premise that Lorentz invariance holds exactly at all scales. Nonetheless, several lines of research aimed at unifying gravity with quantum principles indicate that this symmetry may instead arise as an approximate feature of low--energy physics. In a number of such scenarios, additional geometric ingredients are expected to become relevant near experimentally accessible regimes, potentially leading to small but observable departures from exact invariance \cite{kostelecky1989spontaneous,colladay1997cpt,kostelecky2004gravity,kostelecky1999constraints,kostelecky2011data}. A widely studied route toward this outcome involves spontaneous symmetry breaking triggered by dynamical fields. When a field acquires a nonzero vacuum expectation value, the ground state no longer respects the full Lorentz group and a preferred spacetime direction is dynamically selected.

Bumblebee models constitute a streamlined realization of this mechanism. Rather than introducing explicit symmetry--breaking terms, these constructions rely on a vector field governed by a potential whose minimum enforces a fixed norm. The field relaxes to a configuration of constant magnitude, generating a background that defines an orientation in spacetime. Although the gravitational dynamics remain internally consistent, the presence of this background modifies the geometric structure and alters the propagation and interaction of fields. In this way, the framework provides a systematic setting for investigating deviations from standard relativistic behavior \cite{Bluhm:2019ato,Bluhm:2023kph,Maluf:2014dpa,Maluf:2013nva,bluhm2008spontaneous,bluhm2005spontaneous}.

A number of proposals that go beyond, or reformulate, general relativity suggest that spacetime may accommodate nontrivial vector backgrounds with direct impact on its symmetry structure \cite{kostelecky1989spontaneous,jacobson2004einstein,kostelecky1991photon}. In these approaches, additional fields included in the effective description often settle into vacuum configurations with nonzero expectation values. Once the ground state acquires such a structure, full Lorentz invariance is no longer preserved, since a preferred direction becomes embedded in the vacuum itself \cite{bluhm2005spontaneous,kostelecky2004gravity}. This idea is implemented explicitly in bumblebee constructions. There, a vector field $B_{\mu}$ is introduced together with a potential of the form $V(B_{\mu}B^{\mu}\mp b^{2})$, which constrains its norm \cite{Liu:2022dcn}. The equations of motion drive the system toward the minimum of this potential, where the field attains a constant magnitude. The resulting background configuration selects a spacetime orientation and thereby realizes spontaneous Lorentz violation in a dynamical way \cite{bluhm2008spontaneous,bluhm2005spontaneous}.

Perturbations about this vacuum naturally divide into two sectors. Fluctuations that preserve the fixed--norm condition propagate as massless modes and resemble gauge--type excitations, displaying photonlike characteristics \cite{bluhm2005spontaneous}. By contrast, deviations that alter the norm correspond to massive excitations, their mass originating from the same potential that stabilizes the vacuum state \cite{bluhm2008spontaneous}.

Once the bumblebee mechanism was consistently embedded in curved spacetime, the vacuum expectation value of the vector field became directly intertwined with the gravitational sector, allowing Lorentz symmetry breaking to modify the geometry itself \cite{Bertolami:2005bh}. From that point onward, the subject evolved along multiple interconnected directions rather than a single line of development. Strong--field gravity provided one of the first testing grounds. The static black hole geometry constructed in \cite{Casana:2017jkc} established a reference background in which Lorentz--violating effects could be examined near the horizon. This solution supported analyses of quantum and semiclassical phenomena, including modifications of entanglement properties \cite{Liu:2024wpa} and shifts in particle emission spectra arising from the deformed spacetime structure \cite{AraujoFilho:2025hkm}. Extensions incorporating antisymmetric Kalb--Ramond fields produced additional black hole families where Lorentz violation is built into the metric from the outset \cite{AraujoFilho:2024ctw}.

As the framework matured, alternative geometric formulations were introduced. In particular, the \textit{metric--affine} approach—where metric and connection are treated independently—led to new exact solutions, first in static form \cite{Filho:2022yrk} and later in a rotating, axially symmetric configuration \cite{AraujoFilho:2024ykw}. These constructions opened the door to further generalizations, including non--commutative extensions \cite{AraujoFilho:2025rvn} and parallel developments in antisymmetric tensor sectors such as Kalb--Ramond gravity \cite{AraujoFilho:2025jcu}.

The impact of fixed--norm vector backgrounds has also been explored beyond black holes. Wormhole geometries and their traversability conditions were revisited in Lorentz--violating contexts \cite{Ovgun:2018xys,AraujoFilho:2024iox,Magalhaes:2025lti,Magalhaes:2025nql}, and black--bounce scenarios supported by $\kappa$--essence fields were proposed within the same setting \cite{Pereira:2025xnw}. On larger scales, anisotropic cosmological solutions reminiscent of Kasner expansions \cite{Neves:2022qyb} and anisotropic stellar models \cite{Neves:2024ggn} were obtained, while gravitational wave propagation was shown to depart from the standard general relativistic pattern \cite{Liang:2022hxd,amarilo2024gravitational}. Modifications of the vacuum structure through the inclusion of a cosmological constant generated additional phenomenological consequences \cite{Maluf:2020kgf,Uniyal:2022xnq}.
Furthermore, particle propagation in Lorentz--breaking geometries has received sustained attention. Neutrino deflection and related effects were analyzed in purely metric realizations \cite{Shi:2025plr}, in \textit{metric--affine} versions \cite{Shi:2025ywa}, and in tensorial extensions of the mechanism \cite{Shi:2025rfq}, complemented by further phenomenological studies \cite{Khodadi:2023yiw,Khodadi:2022mzt}.

In recent years, the spectrum of Lorentz--violating black hole solutions has broadened considerably, especially with the construction of geometries associated with distinct vacuum realizations of the bumblebee mechanism \cite{Liu:2025oho,Zhu:2025fiy}. Following the introduction of these configurations, the static sector was analyzed in depth, including its gravitational properties and the constraints imposed on its parameter space \cite{AraujoFilho:2025zaj}. This same background was later employed to investigate neutrino propagation and oscillation phenomena in Lorentz--breaking spacetimes \cite{Shi:2025tvu}. Rotational effects were subsequently incorporated. By applying an improved Newman--Janis algorithm to the static seed, an axisymmetric extension was derived, producing a rotating counterpart \cite{Kumar:2025bim}. The phenomenology of these geometries has also been explored in astrophysical contexts, addressing accretion processes \cite{Shi:2025hfe}, particle production and entanglement behavior \cite{AraujoFilho:2025nmc}, quasinormal spectra \cite{AraujoFilho:2025zaj}, and further aspects of neutrino dynamics \cite{Shi:2025tvu}.

In this work, the main objective is to investigate the Casimir effect associated with a scalar field coupled to gravity, considering as background static and spherically symmetric black hole spacetimes arising in Bumblebee gravity for three different cases, as studied in Refs.~\cite{metric1, metric2, metric31, metric32}. The Casimir effect, originally proposed in 1948 by Hendrik B. G. Casimir~\cite{Casimir}, is a phenomenon inherent to quantum field theory that emerges from modifications of the vacuum fluctuations induced by boundary conditions or nontrivial topology. In its simplest realization, it manifests as an attractive interaction between two parallel conducting plates, resulting from the alteration of the zero-point energy spectrum of the quantum field.

The Casimir effect plays an important role in several areas of physics, including quantum field theory, gravitation and cosmology, condensed matter physics, as well as atomic and molecular physics. It was first experimentally confirmed by Marcus Sparnaay~\cite{Sparnaay}, and subsequent experiments have verified this effect with a high degree of precision~\cite{Exp1, Exp2}. In the present work, the Casimir effect is investigated by taking into account topological contributions arising from spatial compactification, implemented through the topological structure of the Thermo Field Dynamics (TFD) formalism.

TFD is a real-time formalism developed to investigate thermal and finite-size effects in quantum field theory~\cite{thermofield, Book, Book2}. It is based on the equivalence between the statistical average of an operator and its vacuum expectation value in an enlarged Hilbert space. The formalism relies on two fundamental ingredients: the doubling of the Hilbert space and the Bogoliubov transformation. In this framework, a field theory defined on the topology $\Gamma_D^d = (\mathbb{S}^1)^d \times \mathbb{R}^{D-d}$, with $1 \leq d \leq D$, is considered, where $D$ denotes the space-time dimension and $d$ is the number of compactified dimensions. This construction allows any subset of the $\mathbb{R}^D$ manifold to be compactified. The circumference of the $n$-th $\mathbb{S}^1$ is specified by the parameter $\alpha_n$, and the compactification parameters are collectively written as $\alpha = (\alpha_0, \alpha_1, \cdots, \alpha_{D-1})$.

Finite-size effects, such as the Casimir effect, can be incorporated by introducing an appropriate compactification scheme. In particular, the topology $\Gamma_4^1$ with $\alpha = (0, i2d, 0, 0)$ corresponds to a compactification along the radial coordinate $r$, thereby encoding spatial size effects within the formalism. This choice is especially suitable when considering static and spherically symmetric black hole spacetimes as described in~\cite{metric1, metric2, metric31, metric32}.

This paper is organized as follows. In Section \ref{Sec2}, the coupling between a massive scalar field and gravity is introduced, and the energy-momentum tensor together with its vacuum expectation value are calculated. In Section \ref{Sec3}, a brief introduction to the TFD formalism is presented. In Section \ref{Sec4}, three different static and spherically symmetric black hole spacetimes arising in Bumblebee gravity are described. In Section \ref{Sec5}, the results are presented and discussed. The Casimir energy and pressure for this cosmological background are investigated in subsections \ref{SS1} and \ref{SS2}, respectively. Finally, concluding remarks are provided in Section \ref{Sec6}.

%%%%%%%%%%%%%%%%%%%%%%%%%%%%%%%%%%%%%%%%%%%%%%%%%%%%%%%%%%%%%%%%%%%%%%%%%%%%%%%%%%%%%%%%%%%%%%%%%%%%%%%%%%%%%%%%%%%%%%%%%%%%%%%%%%%%%%%%%%%%%%%%%%%%%%%%%%%%%%%%%%%%%%%%%%%%%%%%%%%%%%%%%%%%%%%%%%%%%%%%%%%%%%%%%%%%%%%%%%%%%%%%%%%%%%%%%%%%%%%%%%%%%%%%%%%%%%%%%%%%%%%%%%%%%%%%%%%%%%%%%%%%%%%%%%%%%%%%%%%%%%%%%%%%%%%%%%%%%%%%%%%%%%%%%%%%%%%%%%%%%%%%%%%%%%%%%%%%%%%%%%%%%%%%

\section{Massive scalar field}\label{Sec2}

This part develops the matter sector and specifies the spacetime geometry in which it propagates. We begin by formulating the dynamics of a massive scalar degree of freedom, allowing for a non--minimal coupling to curvature. From this action, the corresponding energy--momentum tensor is derived through variation with respect to the metric, establishing the source structure associated with the field. Once the matter content has been characterized, the gravitational environment is fixed by adopting the bumblebee black hole configuration as the background spacetime. The scalar field is therefore analyzed in a geometry shaped by spontaneous Lorentz symmetry breaking. The starting point is the Lagrangian density describing a massive scalar field with non--minimal curvature coupling, which provides the foundation for all subsequent calculations.
\begin{equation}
    \mathcal{L}=\sqrt{-g}\,\left[\frac{1}{2}\nabla_{\mu}\phi\nabla^{\mu}\phi-\frac{1}{2}m^2\phi^2+f(\phi)R\right],\label{eq02}
\end{equation}
with $g$ denoting the determinant of the metric tensor, $\nabla_\mu$ representing the covariant derivative compatible with the spacetime connection, and $R$ corresponding to the Ricci scalar constructed from that geometry. The function $f(\phi)$ specifies how the scalar field couples non--minimally to curvature.

Within the Thermo Field Dynamics framework, the central object required for physical applications is the energy--momentum tensor, which is defined as
\begin{eqnarray}
    T_{\mu\nu}=-\frac{2}{\sqrt{-g}}\frac{\delta S}{\delta g^{\mu\nu}}.
\end{eqnarray}
Here, $S$ denotes the action associated with the model introduced in Eq.~\eqref{eq02}. For a scalar field, the corresponding energy--momentum tensor takes the general form
\begin{equation}
    T_{\mu\nu}=\frac{1}{2}g_{\mu\nu}\partial^{\kappa}\phi\partial_{\kappa}\phi-\partial_{\mu}\phi\partial_{\nu}\phi-\frac{1}{2}g_{\mu\nu}m^2\phi^2-2\left(R_{\mu\nu}-\frac{1}{2}g_{\mu\nu}\,R+g_{\mu\nu}\square-\nabla_{\mu}\nabla_{\nu}\right)f(\phi).
\end{equation}
Notice that if we specifies the coupling function as $f(\phi)=\tfrac{1}{2}\,\xi \phi^{2}$, with $\xi$ denoting the conformal coupling constant, the resulting expression for the energy--momentum tensor becomes
\ie
T_{\mu\nu}=\frac{1}{2}g_{\mu\nu}g^{\kappa\lambda}\partial_{\kappa}\phi\partial_{\lambda}\phi-\partial_{\mu}\phi\partial_{\nu}\phi-\left[\xi\left(R_{\mu\nu}-\frac{1}{2}g_{\mu\nu}\,R+g_{\mu\nu}\square-\nabla_{\mu}\nabla_{\nu}\right)\textcolor{black}{+\frac{1}{2}g_{\mu\nu}m^2}\right]\phi^2.
\fe

In order to regularize the short--distance singularities that arise from products of field operators evaluated at the same spacetime point, the tensor is instead defined in a point–separated form, namely
\begin{align}
    T_{\mu\nu}(x)=&\lim_{x'\to x}\tau \Bigl\{\frac{1}{2}g_{\mu\nu}\partial^{\kappa}\phi(x)\partial'_{\kappa}\phi(x')-\partial_{\mu}\phi(x)\partial'_{\nu}\phi(x')
    \nonumber\\&-\left[\xi\left(R_{\mu\nu}-\frac{1}{2}g_{\mu\nu}\,R+g_{\mu\nu}\square-\nabla_{\mu}\nabla_{\nu}\right)+\frac{1}{2}g_{\mu\nu}m^2\right]\phi(x)\phi(x')\Bigr\},
\end{align}
with $\tau$ denotes the time--ordering operator. Under canonical quantization, the fields satisfy the equal time commutation relation given by
\begin{equation}
    [\phi(x), \partial'^{\mu}\phi(x')]=i\,n^{\mu}_{0}\,\delta(\vec{x}-\vec{x'}),
\end{equation}
and 
\begin{equation}
    \partial^{\rho}\,\theta(x^0-x'^0)=i\,n_0^\mu\,\delta(x^0-x'^0),
\end{equation}
with $n_0^\mu=(1,0,0,0)$ denoting a unit timelike vector, $\theta(x_0-x_0')$ the Heaviside step function, and $\delta(\vec{x}-\vec{x}')$ the standard three--dimensional Dirac delta. For brevity, we use the shorthand $x^\mu\equiv x=(x^0,\vec{x})$. Under these definitions, the energy--momentum tensor can be rewritten as
\begin{equation}
    T_{\mu\nu}(x)=\lim_{x'\to x}\left\{\Gamma_{\mu\nu}\,\tau\left(\phi(x)\phi(x')\right)-I_{\mu\nu}\delta (x-x')\right\},
\end{equation}
in which
\begin{equation}
 \Gamma_{\mu\nu}=\frac{1}{2}g_{\mu\nu}\partial^{\kappa}\partial'_{\kappa} -\partial_{\mu}\partial'_{\nu}-\xi\left(R_{\mu\nu}-\frac{1}{2}g_{\mu\nu}\,R+g_{\mu\nu}\square-\nabla_{\mu}\nabla_{\nu}\right)-\frac{1}{2}g_{\mu\nu}m^2  \label{8}
\end{equation}
and
\begin{equation}
    I_{\mu\nu}=-\frac{i}{2}\,g_{\mu\nu}n_{0}^{\kappa}n_{0\kappa}+i\,n_{0\mu}n_{0\nu}.
\end{equation}

To characterize both finite–temperature contributions and finite–size corrections, one evaluates the vacuum expectation value of the energy–momentum tensor, written as
\begin{eqnarray}\label{Tavr}
   \langle T_{\mu\nu}(x)\rangle &=&\bra{0}T_{\mu\nu}(x)\ket{0}\nonumber\\
   &=&\lim_{x'\to x}\left[i\,\Gamma_{\mu\nu}\, G_0(x-x')-I_{\mu\nu}\,\delta(x-x')\right],
\end{eqnarray}
where
\begin{equation}
    i\,G_0(x-x')=\bra{0}\tau\left(\phi(x)\phi(x')\right)\ket{0},
\end{equation}
is, therefore, the massive scalar propagator, which admits the representation given by
\begin{equation}\label{Gzerox}
    G_{0}(x-x')=-\frac{i\,m}{4\pi^2}\frac{K_{1}\left(m\sqrt{-(x-x')^2}\right)}{\sqrt{-(x-x')^2}}.
\end{equation}
In coordinate space, the propagator is expressed as a function of $x$. Equivalently, the same quantity can be represented in momentum space as a function of $k$, namely: 
\begin{equation}\label{kpropagator}
    G_0(k)=\frac{1}{k^2-m^2+i\epsilon}.
\end{equation}

In the next section, the necessary tools from the TFD formalism are introduced in order to incorporate finite-size effects into the Green function. Furthermore, the vacuum expectation value of the energy--momentum tensor is evaluated within this framework.

%%%%%%%%%%%%%%%%%%%%%%%%%%%%%%%%%%%%%%%%%%%%%%%%%%%%%%%%%%%%%%%%%%%%%%%%%%%%%%%%%%%%%%%%%%%%%%%%%%%%%%%%%%%%%%%%%%%%%%%%%%%%%%%%%%%%%%%%%%%%%%%%%%%%%%%%%%%%%%%%%%%%%%%%%%%%%%%%%%%%%%%%%%%%%%%%%%%%%%%%%%%%%%%%%%%%%%%%%%%%%%%%%%%%%%%%%%%%%%%%%%%%%%%%%%%%%%%%%%%%%%%%%%%%%%%%%%%%%%%%%%%%%%%%%%%%%%%%%%%%%%%%%%%%%%%%%%%%%%%%%%%%%%%%%%%%%%%%%%%%%%%%%%%%%%%%%%%%%%%%%%%%%%%%%%%%%%%%%%%%%%%%%%%%%%%%%%%%%%%%%%%%%%%%%%%%%%

\section{Thermo Field Dynamics}\label{Sec3}

This section develops a finite–size quantum field theory using real–time techniques within the Thermo Field Dynamics (TFD) framework~\cite{thermofield, Book, Book2}. In this setting, spatial compactification can be implemented directly at the level of Green functions, which makes it suitable for describing Casimir--type configurations. In particular, the radial coordinate is compactified, and the effect of a finite extension is encoded through a modification of the coordinate space propagator. Introducing a compactification scale $d$, the propagator is accordingly redefined as
\begin{equation}
    G^{(11)}(x-x^\prime;d)=i\int\frac{\mathrm{d}^4q}{(2\pi)^4}e^{-iq(x-x^\prime)}G^{(11)}(q;d),
\end{equation}
with $G^{(11)}(q;d)$ being the momentum space propagator modified by the finite extension, corresponding to the $(11)$ component in the TFD formalism and incorporating the compactification scale $d$, which is written as
\begin{equation}
    G^{(11)}(q;d)=G_0(q)+\sum_{j=1}^{\infty} e^{-2idq_3}\left[G^*_0(q)-G_0(q)\right].
\end{equation}
In this notation, $G_0(q)$ represents the standard Green function introduced in Eq.~\eqref{kpropagator}, while $G_0^*(q)$ denotes its complex conjugate. The above structure matches the real time treatment of fields under compactification. A detailed derivation of this result is presented in Ref.~\cite{Book}.

The parameter $d$ is identified with the plate separation of a spherical capacitor, i.e., the radial gap between two concentric spherical shells. With this identification, Casimir–type effects are modeled by restricting the scalar field to a finite radial domain and by enforcing the corresponding boundary conditions at the shells.

With the compactified propagator in hand, Eq.~(\ref{Tavr}) can be recast entirely in terms of finite--size quantities. One then writes the vacuum expectation value of the energy--momentum tensor as
\begin{eqnarray}
   \langle T_{\mu\nu}^{(11)}(x;d)\rangle =\lim_{x'\to x}\left[i\,\Gamma_{\mu\nu}\, G_0^{(ab)}(x-x';d)-I_{\mu\nu}\,\delta(x-x')\delta^{ab}\right].
\end{eqnarray}
In order to extract physically meaningful and finite results, the energy–momentum tensor must be renormalized. This is achieved by implementing the Casimir subtraction scheme, in which the unbounded (free) contribution is removed from the compactified expression. The finite, renormalized energy--momentum tensor is then written as
\ie
{\cal T}_{\mu\nu }^{(11)}(x;d)=\langle T_{\mu\nu}^{(11)}(x;d)\rangle-\langle T_{\mu\nu}^{(11)}(x)\rangle.
\fe
After performing this subtraction, the resulting expression corresponds to an observable, finite quantity. In explicit form,
\ie
{\cal T}_{\mu\nu }^{(11)}(x;d)=i\lim_{x'\rightarrow x}\left\{\Gamma_{\mu\nu}(x,x')\overline{G}_0^{(11)}(x-x';d)\right\},\label{VEV}
\fe
with 
\ie
\overline{G}_0^{(11)}(x-x';d)=\sum_{j=1}^{\infty}\left[G_0^{*}(x-x'-2jd\hat{e}_r)-G_0(x-x^\prime-2jd\hat{e}_r)\right].
\fe

From the expressions obtained above, one immediately arrives at
\begin{eqnarray}
\mathcal{T}_{\mu\nu}^{(11)}=2i\sum_{j=1}^{\infty}\lim_{x'\rightarrow x}\left\{\Gamma_{\mu\nu}(x,x')G_0(x-x'-2jd\hat{e}_r)\right\},\label{eq01}
\end{eqnarray}
from which we can extract the relevant observables, in particular the Casimir energy and the associated pressure in the bumblebee black hole backgrounds, as discussed in the next section.

%%%%%%%%%%%%%%%%%%%%%%%%%%%%%%%%%%%%%%%%%%%%%%%%%%%%%%%%%%%%%%%%%%%%%%%%%%%%%%%%%%%%%%%%%%%%%%%%%%%%%%%%%%%%%%%%%%%%%%%%%%%%%%%%%%%%%%%%%%%%%%%%%%%%%%%%%%%%%%%%%%%%%%%%%%%%%%%%%%%%%%%%%%%%%%%%%%%%%%%%%%%%%%%%%%%%%%%%%%%%%%%%%%%%%%%%%%%%%%%%%%%%%%%%%%%%%%%%%%%%%%%%%%%%%%%%%%%%%%%%%%%%%%%%%%%%%%%%%%%%%%%%%%%%%%%%%%%%%%%%%%%%%%%%%%%

\section{The bumblebee solutions}\label{Sec4}

The analysis is organized into three separate scenarios, each defined by a different spacetime geometry. We begin with the first configuration, corresponding to the bumblebee black hole obtained within the \textit{metric} formulation \cite{metric1}, whose line element is given by
\begin{align}
\mathrm{d}s^{2} = - \left(1 - \dfrac{2M}{r}\right)\mathrm{d}t^{2} + (1+\ell)\left(1 - \dfrac{2M}{r} \right)^{-1} \mathrm{d}r^{2} + r^{2}\mathrm{d}\Omega^{2}.\label{sol1}
\end{align}

The second configuration corresponds to the bumblebee solution derived within the \textit{metric--affine} framework \cite{metric2}, whose spacetime structure is expressed as
\begin{equation}
\begin{split}
  \mathrm{d}s^2= &- \frac{\left(1-\frac{2M}{r}\right)\mathrm{d}t^2}{\sqrt{\left(1+\frac{3X}{4}\right)\left(1-\frac{X}{4}\right)}}+\frac{\mathrm{d}r^2}{\left(1-\frac{2M}{r}\right)}\sqrt{\frac{\left(1+\frac{3X}{4}\right)}{\left(1-\frac{X}{4}\right)^3}}\\
  & +r^{2}\left(\mathrm{d}\theta^2 +\sin^{2}{\theta}\mathrm{d}\phi^2\right).\label{sol2}
\end{split}
\end{equation}

The third configuration, following the same reasoning, corresponds to the bumblebee black hole constructed in Refs.~\cite{metric31, metric32}, whose line element is written as
\begin{align}
\mathrm{d}s^{2} = - \dfrac{1}{1+\chi}\left(1 - \dfrac{2M}{r}\right)\mathrm{d}t^{2} + (1+\chi)\left(1 - \dfrac{2M}{r} \right)^{-1} \mathrm{d}r^{2} + r^{2}\mathrm{d}\Omega^{2}.\label{sol3}
\end{align}
The parameters $\chi$, $X$, and $\ell$ are constant quantities fixed by the corresponding vacuum expectation values of the bumblebee sector.

\begin{figure}[ht]
    \centering
    \includegraphics[width=1\linewidth]{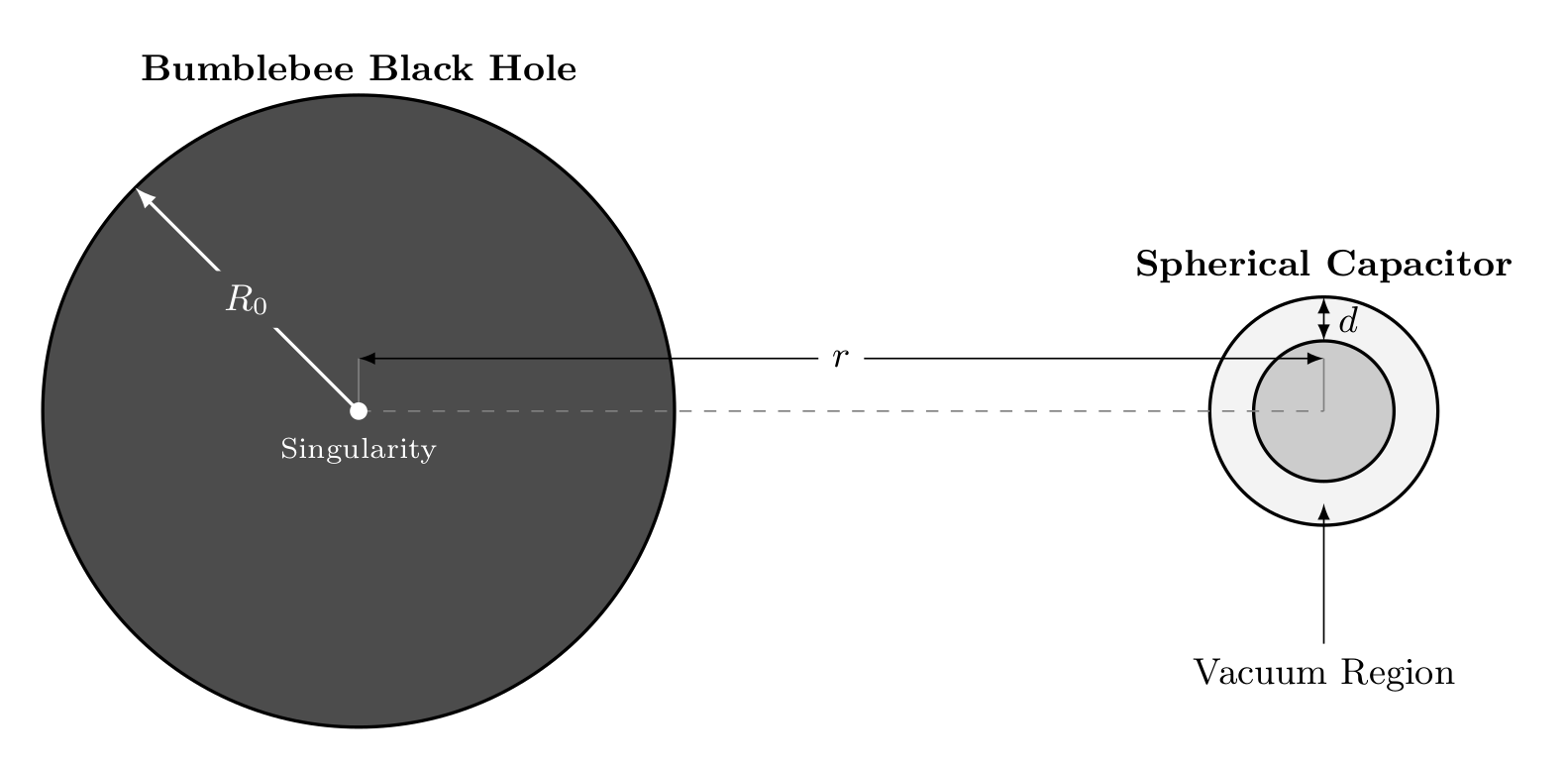}
    \caption{Spherical capacitor placed at radial coordinate $r$ from the center of a bumblebee black hole of radius $R_0$. The two concentric shells are separated by a distance $d$. The Casimir effect manifests as an attractive force between the shells, generated by vacuum fluctuations of the confined scalar field modes within the intervening region.}
    \label{fig1}
\end{figure}

Each of the three geometries represents a distinct black hole configuration, differing solely in the structure assumed by the bumblebee field. By implementing the TFD framework with spatial compactification, boundary--induced effects can be incorporated into the analysis. This procedure makes it possible to evaluate Casimir--type contributions and to extract observables such as the energy density and pressure directly from the renormalized energy--momentum tensor. A schematic representation of the configuration is displayed in Fig.~\ref{fig1}.
The finite--size corrections will now be examined independently for each metric, highlighting their specific features and common aspects, followed by a comparative visualization of the outcomes.

%%%%%%%%%%%%%%%%%%%%%%%%%%%%%%%%%%%%%%%%%%%%%%%%%%%%%%%%%%%%%%%%%%%%%%%%%%%%%%%%%%%%%%%%%%%%%%%%%%%%%%%%%%%%%%%%%%%%%%%%%%%%%%%%%%%%%%%%%%%%%%%%%%%%%%%%%%%%%%%%%%%%%%%%%%%%%%%%%%%%%%%%%%%%%%%%%%%%%%%%%%%%%%%%%%%%%%%%%%%%%%%%%%%%%%%%%%%%%%%%%%%%%%%%%%%%%%%%%%%%%%%%%%%%%%%%%%%%%%%%%%%%%%%%%%%%%%%%%%%%%%%%%%%%%%%%%%%%%%%%%%%%%%%%%%%%%%%%%%%%%%%%%%%%%%%%%%%%%%%%%%%%%%%%%%%%%%%%%%%%%%%%%%%%%%%%%%%%%%%%%%%%%%%%%%%%%%

\subsection{The first solution}

Substituting the metric solution given in Eq.~\eqref{sol1} into the energy--momentum tensor of Eq.~\eqref{eq01}, and identifying the energy density as $E_1(r,d)=\mathcal{T}_{00}^{11}$, one derives the corresponding Casimir energy. In the massless limit of the scalar field ($m=0$), the result reduces to
\begin{align}
E_1(r,d) &= -\frac{r-R_0}{1440 \pi^2 d^4 (\ell+1)^2 r^5} \biggl\{ \pi^4 (\xi +1) r^2 (r-R_0)^2  + 90 d r (r-R_0) \zeta (3) \left[4 \xi r + (\xi +1) R_0\right] \nonumber \\
&\quad - 15 \pi^2 \xi d^2 \left[2 \ell r (r-R_0) - R_0^2\right] \biggr\}.
\label{eqE1}\end{align}
It is important to emphasize that the massless limit for the scalar field is assumed in order to obtain closed-form expressions for the Casimir energy. This simplifying assumption makes the analytical treatment more tractable and allows for a clearer interpretation of the effects induced by the gravitational background and spatial compactification on the vacuum fluctuations of the field.

In the same manner, the Casimir pressure is obtained by identifying $P_1(r,d)=\mathcal{T}_{11}^{11}$. This leads to
\begin{align}
P_1(r,d) &= -\frac{1}{1440 \pi^2 d^4 (\ell+1) r^4 (r-R_0)} \biggl\{ \pi^4 r^2 (r-R_0)^2 \left[ (2 \xi + 3) r + \xi R_0 \right] \nonumber \\
&\quad + 90 d r R_0 (r-R_0) \zeta (3) \left[ (6 \xi + 1) r - \xi R_0 \right] \nonumber \\
&\quad - 15 \pi^2 \xi d^2 \left[ R_0 \left(R_0^2 - 6 r R_0 + 4 r^2\right) - 2 \ell r^2 (r-R_0) \right] \biggr\}.
\label{eqP1}\end{align}

In both expressions, the horizon scale is fixed by $R_0=2M$, which we identify as the bumblebee radius. The resulting Casimir energy and pressure depend explicitly on the radial location $r$ of the capacitor and on the separation $d$ between the shells. A direct inspection of their signs shows that the interaction may switch between attractive and repulsive regimes, depending on how far the system is placed from the black hole center.

%%%%%%%%%%%%%%%%%%%%%%%%%%%%%%%%%%%%%%%%%%%%%%%%%%%%%%%%%%%%%%%%%%%%%%%%%%%%%%%%%%%%%%%%%%%%%%%%%%%%%%%%%%%%%%%%%%%%%%%%%%%%%%%%%%%%%%%%%%%%%%%%%%%%%%%%%%%%%%%%%%%%%%%%%%%%%%%%%%%%%%%%%%%%%%%%%%%%%%%%%%%%%%%%%%%%%%%%%%%%%%%%%%%%%%%%%%%%%%%%%%%%%%%%%%%%%%%%%%%%%%%%%%%%%%%%%%%%%%%%%%%%%%%%%%%%%%%%%%%%%%%%%%%%%%%%%%%%%%%%%%%%%%%%%%%%%%%%%%%%%%%%%%%%%%%%%%%%%%%%%%%%%%%%%%%%%%%%%%%%%%%%%%%%%%%%%%%%%%%%%%%%%%%%%%%%%%

\subsection{The Second solution}

Following the same procedure adopted before, but now using the geometry specified by Eq.~\eqref{sol2}, we arrive at
\begin{align} 
E_2(r,d) &= -\frac{(r-R_0)(4-X)^{5/2}}{5760 \pi^2 d^4 r^5 (3X+4)^{3/2}} \biggl\{ \pi^4 (\xi +1) r^2 (r-R_0)^2  + 90 d r (r-R_0) \zeta (3) \left[4 \xi r + (\xi +1) R_0\right] \nonumber \\
&\quad + 15 \pi^2 \xi d^2 \left[ (r-R_0)^2 + r^2 - 8 r (r-R_0) \sqrt{\frac{3X+4}{(4-X)^3}} \right] \Biggr\},
\label{eqE2}\end{align}
for the energy expression, and
{\small\begin{align}
P_2(r,d) &= \frac{4-X}{5760 \pi^2 d^4 r^4 (r-R_0) (3X+4)} \biggl\{ \pi^4 r^2 (r-R_0)^2 \left[ 4\xi(r-R_0) - 3(\xi+1)r \sqrt{-3X^2+8X+16} \right] \nonumber \\
&\quad - 90 d r R_0 (r-R_0) \zeta (3) \left[ 4\xi(r-R_0) + (5\xi+1)r \sqrt{-3X^2+8X+16} \right] \nonumber \\
&\quad + 15 \pi^2 \xi d^2 \left[ r (2r^2+2r R_0 - 5R_0^2)\sqrt{-3X^2+8X+16} - 4(r-R_0)\left(R_0^2 + \frac{2r^2(3X+4)}{4-X}\right) \right] \biggr\}
\label{eqP2}\end{align}}
for the Casimir pressure.

In this configuration, the observables display a markedly different dependence on the parameter $X$ when compared with the previous case (with $\ell$). The geometry is now controlled by the constant $X$, which leads to a modified pole structure in the corresponding expressions. Moreover, unlike the first solution, the dynamical corrections affect simultaneously the temporal and radial components of the metric, producing a distinct modification of the spacetime structure.

%%%%%%%%%%%%%%%%%%%%%%%%%%%%%%%%%%%%%%%%%%%%%%%%%%%%%%%%%%%%%%%%%%%%%%%%%%%%%%%%%%%%%%%%%%%%%%%%%%%%%%%%%%%%%%%%%%%%%%%%%%%%%%%%%%%%%%%%%%%%%%%%%%%%%%%%%%%%%%%%%%%%%%%%%%%%%%%%%%%%%%%%%%%%%%%%%%%%%%%%%%%%%%%%%%%%%%%%%%%%%%%%%%%%%%%%%%%%%%%%%%%%%%%%%%%%%%%%%%%%%%%%%%%%%%%%%%%%%%%%%%%%%%%%%%%%%%%%%%%%%%%%%%%%%%%%%%%%%%%%%%%%%%%%%%%%%%%%%%%%%%%%%%%%%%%%%%%%%%%%%%%%%%%%%%%%%%%%%%%%%%%%%%%%%%%%%%%%%%%%%%%%%%%%%%%%%%

\subsection{The Third solution}

Adopting the geometry specified in Eq.~\eqref{sol3} and repeating the same computational steps, one finds that the corresponding Casimir energy takes the form
\begin{align}
E_3(r,d) &= -\frac{r-R_0}{1440 \pi^2 d^4 (\chi +1)^3 r^5} \biggl\{ \pi^4 (\xi +1) r^2 (r-R_0)^2 + 90 d r (r-R_0) \zeta (3) \left[4 \xi r + (\xi +1) R_0\right] \nonumber \\
&\quad - 15 \pi^2 \xi d^2 \left[2 \chi r (r-R_0) - R_0^2\right] \biggr\},
\label{eqE3}\end{align}
and the pressure is expressed as
\begin{align}
P_3(r,d) &= -\frac{1}{1440 \pi^2 d^4 r^4 (r-R_0) (\chi +1)^2} \biggl\{ \pi^4 r^2 (r-R_0)^2 \left[ r \left(3 \chi (\xi +1)+2 \xi +3\right) + \xi R_0 \right] \nonumber \\
&\quad + 90 d r R_0 (r-R_0) \zeta (3) \left[ r \left(\xi (5 \chi +6)+\chi +1\right) - \xi R_0 \right] \nonumber \\
&\quad + 15 \pi^2 \xi d^2 \left[ 2 \chi (\chi +1) r^3 - 2 (\chi +1) (\chi +2) r^2 R_0 + (5 \chi +6) r R_0^2 - R_0^3 \right] \biggr\}.
\label{eqP3}\end{align}

At first sight, these expressions may seem close to those obtained for the case in Eq.~\eqref{sol1}, since the correction to the $g_{11}$ component has the same structure. However, the extra deformation in the temporal sector alters the resulting observables in a nontrivial way. With all three cases in place, the physical consequences can be examined in a unified manner and the outcomes compared across the different geometries.

%%%%%%%%%%%%%%%%%%%%%%%%%%%%%%%%%%%%%%%%%%%%%%%%%%%%%%%%%%%%%%%%%%%%%%%%%%%%%%%%%%%%%%%%%%%%%%%%%%%%%%%%%%%%%%%%%%%%%%%%%%%%%%%%%%%%%%%%%%%%%%%%%%%%%%%%%%%%%%%%%%%%%%%%%%%%%%%%%%%%%%%%%%%%%%%%%%%%%%%%%%%%%%%%%%%%%%%%%%%%%%%%%%%%%%%%%%%%%%%%%%%%%%%%%%%%%%%%%%%%%%%%%%%%%%%%%%%%%%%%%%%%%%%%%%%%%%%%%%%%%%%%%%%%%%%%%%%%%%%%%%%%%%%%%%%%%%%%%%%%%%%%%%%%%%%%%%%%%%%%%%%%%%%%%%%%%%%%%%%%%%%%%%%%%%%%%%%%%%%%%%%%%%%%%%%%%%

\section{Results and Discussions}\label{Sec5}

In this section, the expressions derived above are discussed in more detail by inspecting the plots and analyzing the obtained results.

%%%%%%%%%%%%%%%%%%%%%%%%%%%%%%%%%%%%%%%%%%%%%%%%%%%%%%%%%%%%%%%%%%%%%%%%%%%%%%%%%%%%%%%%%%%%%%%%%%%%%%%%%%%%%%%%%%%%%%%%%%%%%%%%%%%%%%%%%%%%%%%%%%%%%%%%%%%%%%%%%%%%%%%%%%%%%%%%%%%%%%%%%%%%%%%%%%%%%%%%%%%%%%%%%%%%%%%%%%%%%%%%%%%%%%%%%%%%%%%%%%%%%%%%%%%%%%%%%%%%%%%%%%%%%%%%%%%%%%%%%%%%%%%%%%%%%%%%%%%%%%%%%%%%%%%%%%%%%%%%%%%%%%%%%%%%%%%%%%%%%%%%%%%%%%%%%%%%%%%%%%%%%%%%%%%%%%%%%%%%%%%%%%%%%%%%%%%%%%%%%%%%%%%%%%%%%%

\subsection{The energy}\label{SS1}

\begin{figure}
    \centering
    \includegraphics[scale=0.4]{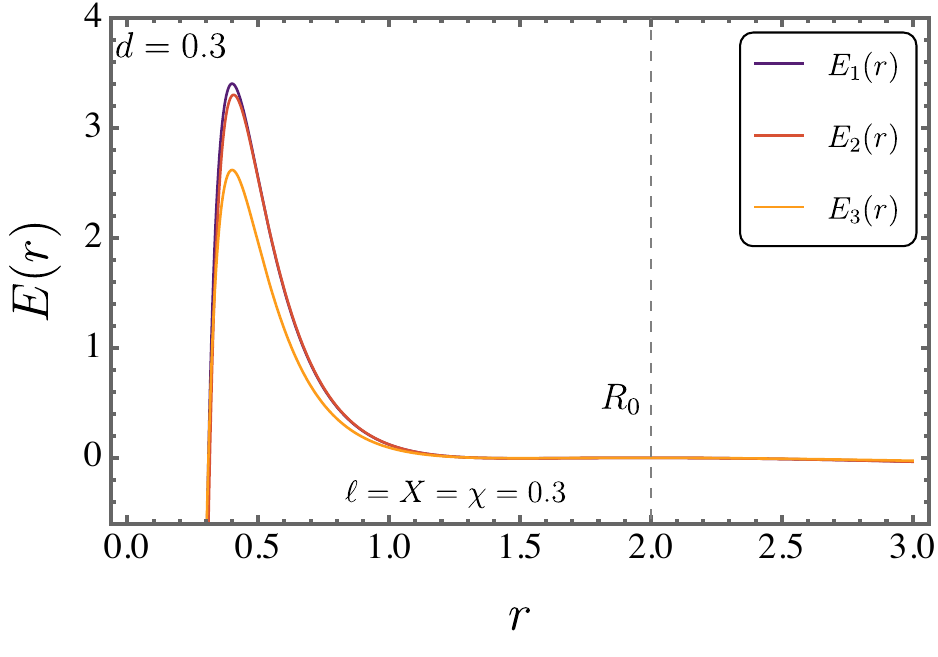}
    \includegraphics[scale=0.44]{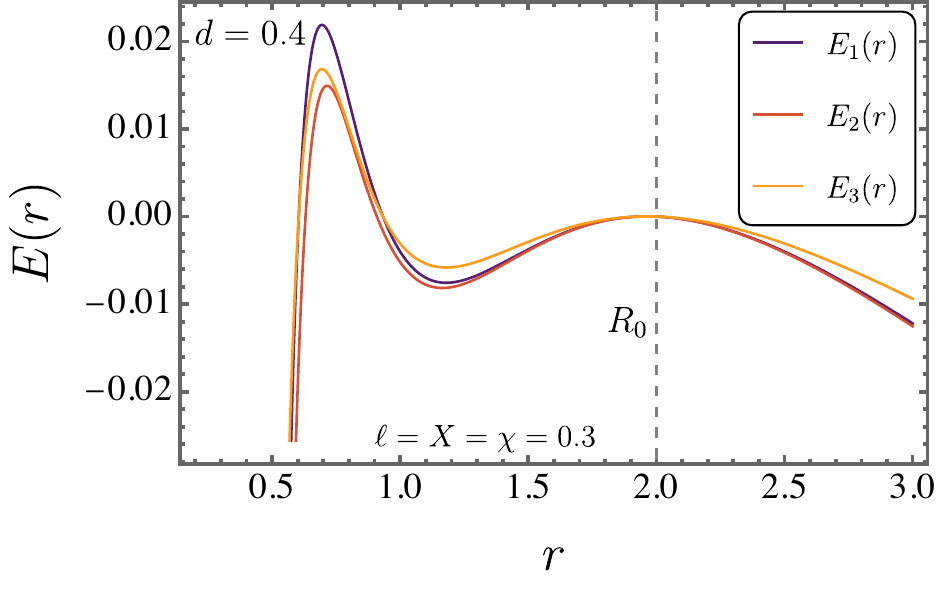}
    \includegraphics[scale=0.44]{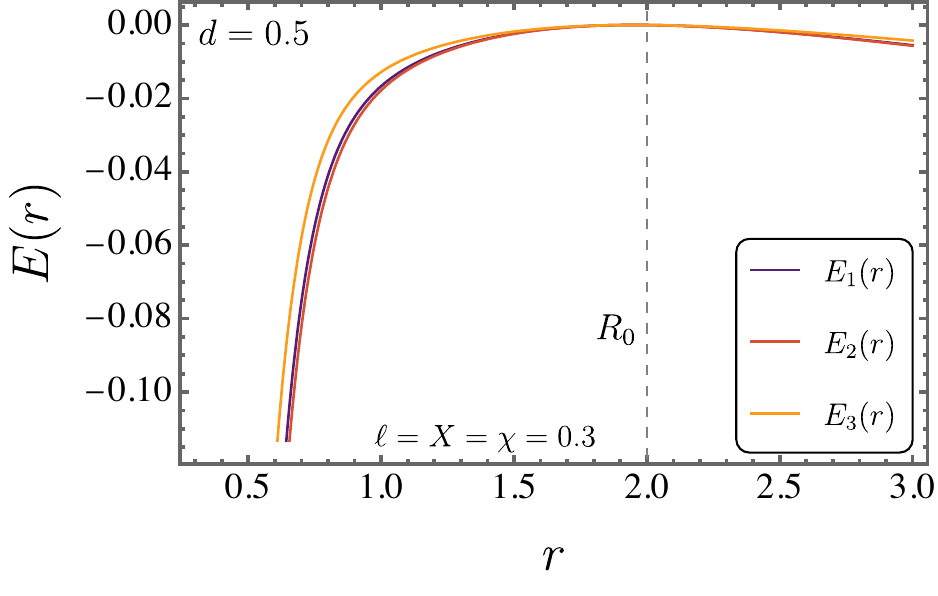}
    \includegraphics[scale=0.44]{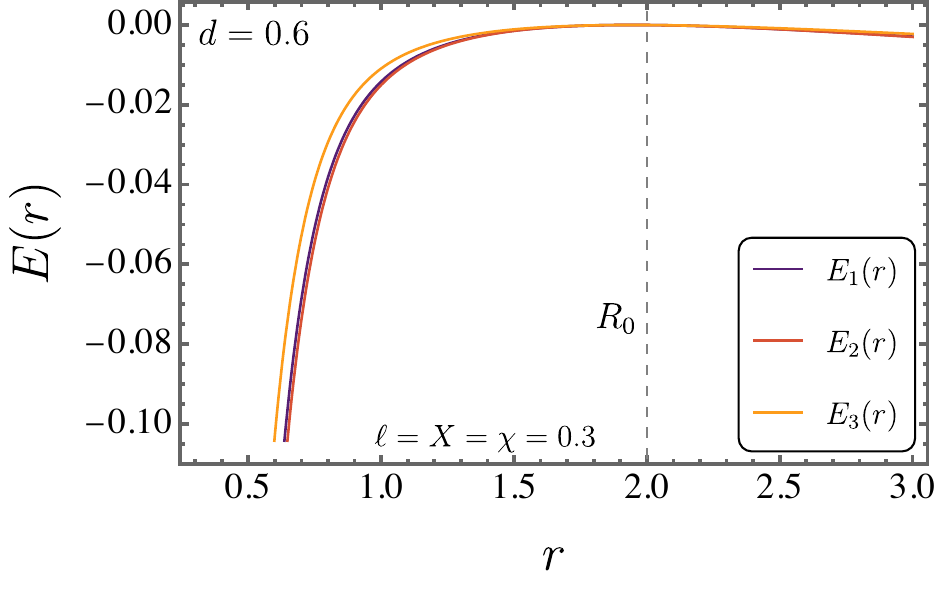}
    \caption{Casimir energy $E_i(r)$ as a function of the radial coordinate $r$ for the three bumblebee black hole geometries defined in Eqs.~\eqref{eqE1}--\eqref{eqE3}. The parameters are fixed to $M=1$, $\xi=1/4$, and $\ell=X=\chi=0.3$. The curves correspond to plate separations $d=0.3$, $0.4$, $0.5$, and $0.6$. }
    \label{figenergy1}
\end{figure}

\begin{figure}
    \centering
    \includegraphics[scale=0.44]{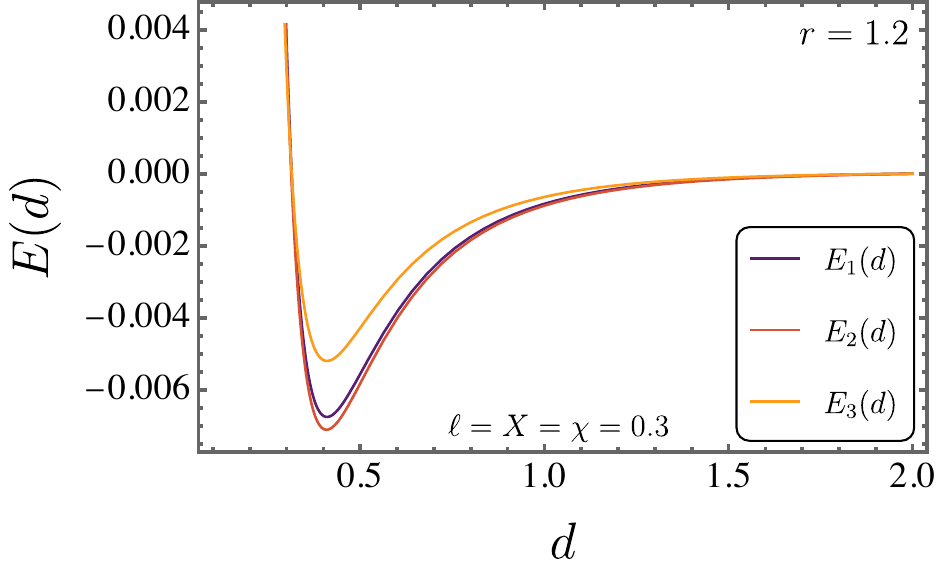}
    \includegraphics[scale=0.44]{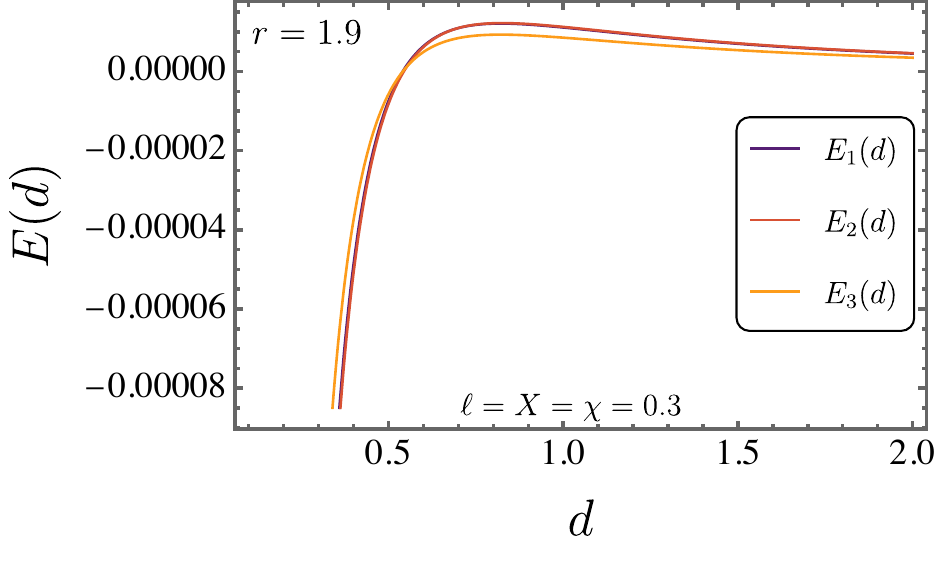}
    \includegraphics[scale=0.44]{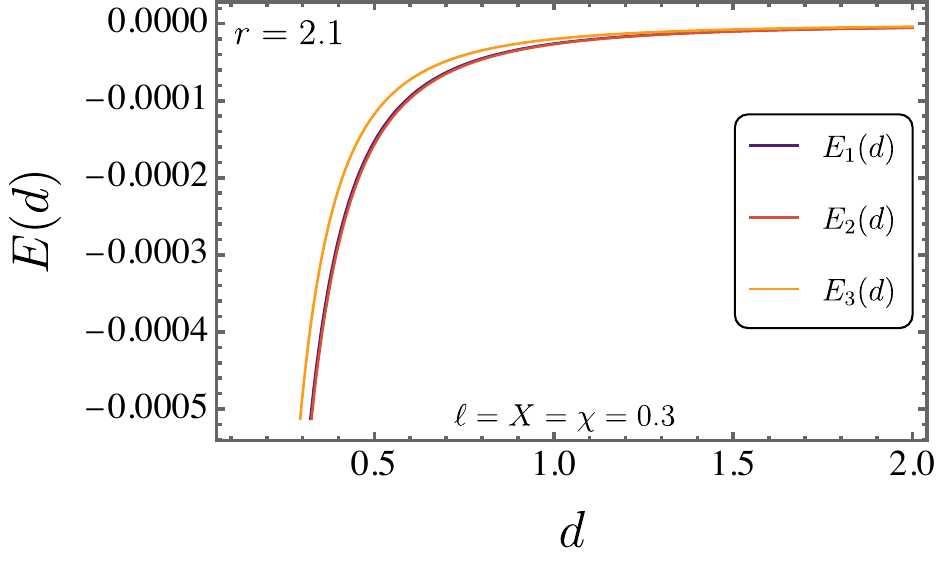}
    \includegraphics[scale=0.44]{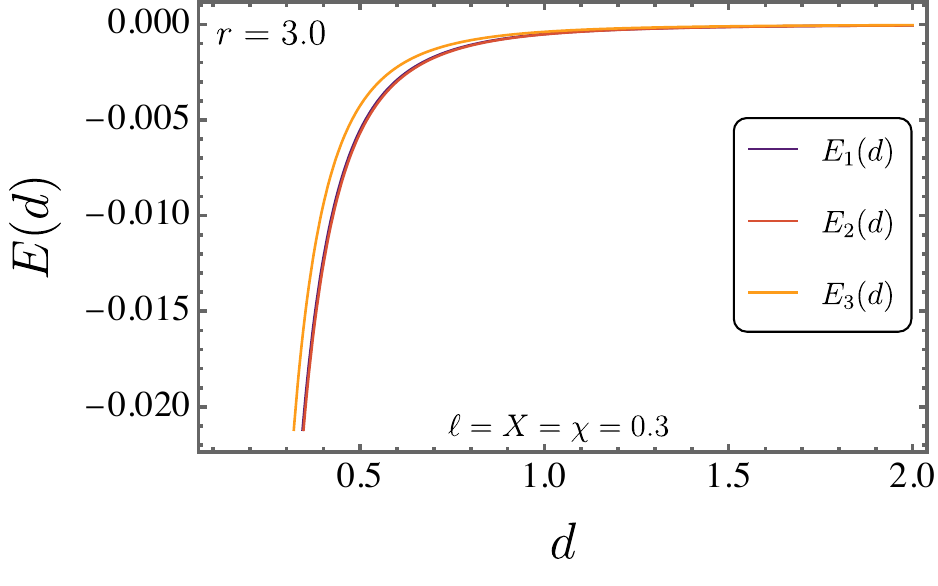}
    \caption{Casimir energy $E_i(d)$ as a function of the plate separation $d$ for fixed radial positions, with $M=1$, $\xi=1/4$, and $\ell=X=\chi=0.3$}
    \label{figenergy2}
\end{figure}

Figs. \ref{figenergy1} and \ref{figenergy2} display the Casimir energies $E_i(r,d)$ corresponding to the three bumblebee geometries obtained in Eqs.~\eqref{eqE1}--\eqref{eqE3}. In all panels we fix $M=1$, $\xi=1/4$, and choose identical Lorentz–violating parameters $\ell=X=\chi=0.3$ in order to allow a direct comparison between the three backgrounds. The radial profiles are shown for plate separations $d=0.3$, $0.4$, $0.5$, and $0.6$. For $r\gg R_0$, the three energies converge toward the flat–space expression $E=-\pi^2 f/(1440 d^4)$ \cite{ThermalBH, ThermalBH2}, differing only by model–dependent factors. As $r\to R_0$, all configurations exhibit a smooth suppression of the energy, which vanishes at the Bumblebee radius. The main differences emerge in the interior region.

Although $E_1$, $E_2$, and $E_3$ share the same qualitative structure, their magnitudes and radial gradients differ noticeably. The first configuration shows the mildest deformation relative to the flat–space behavior. The second case produces stronger deviations near the horizon, reflecting its distinct pole structure. The third geometry yields the largest departure in the interior region, indicating that simultaneous modifications of the temporal and radial metric components enhance the sensitivity of the vacuum energy to the background.

Varying the plate separation confirms that smaller values of $d$ amplify the magnitude of the energy in all three models, while larger separations soften the interaction. However, the relative spacing between $E_1$, $E_2$, and $E_3$ remains controlled primarily by the geometric structure of each solution rather than by $d$ alone.

For a fixed radial position inside the horizon ($r<R_{0}$), and in particular for $r=1.2$ in the configuration considered, the dependence on the plate separation $d$ reveals a well–defined hierarchy among the three energies. In this region one finds
$E_{2}(d) < E_{1}(d) < E_{3}(d)$.
Since the Casimir interaction is attractive when the energy is negative, this ordering implies that the strongest attractive interaction (largest magnitude) is produced by the second configuration, while the third case yields comparatively weaker attraction. Therefore, the geometry associated with non–metricity enhances the magnitude of the vacuum energy inside the horizon relative to the purely radial deformation.
On the other hand, in the exterior region ($r>R_{0}$) the ordering changes to
$E_{3}(d) > E_{1}(d) > E_{2}(d)$.

Although all three solutions approach the flat space limit asymptotically, the third configuration consistently produces the largest energy values outside the horizon. This behavior reflects the fact that the third solution incorporates the most general vacuum structure for the bumblebee field, with simultaneous modifications in both temporal and radial sectors of the metric. Consequently, its influence on the vacuum fluctuations remains more pronounced even in the weak–field regime.

Another quantity of significant physical interest related to the Casimir energy is the pressure, which we discuss below.

%%%%%%%%%%%%%%%%%%%%%%%%%%%%%%%%%%%%%%%%%%%%%%%%%%%%%%%%%%%%%%%%%%%%%%%%%%%%%%%%%%%%%%%%%%%%%%%%%%%%%%%%%%%%%%%%%%%%%%%%%%%%%%%%%%%%%%%%%%%%%%%%%%%%%%%%%%%%%%%%%%%%%%%%%%%%%%%%%%%%%%%%%%%%%%%%%%%%%%%%%%%%%%%%%%%%%%%%%%%%%%%%%%%%%%%%%%%%%%%%%%%%%%%%%%%%%%%%%%%%%%%%%%%%%%%%%%%%%%%%%%%%%%%%%%%%%%%%%%%%%%%%%%%%%%%%%%%%%%%%%%%%%%%%%%%%%%%%%%%%%%%%%%%%%%%%%%%%%%%%%%%%%%%%%%%%%%%%%%%%%%%%%%%%%%%%%%%%%%%%%%%%%%%%%%%%%%

\subsection{The pressure}\label{SS2}

\begin{figure}
    \centering
    \includegraphics[scale=0.44]{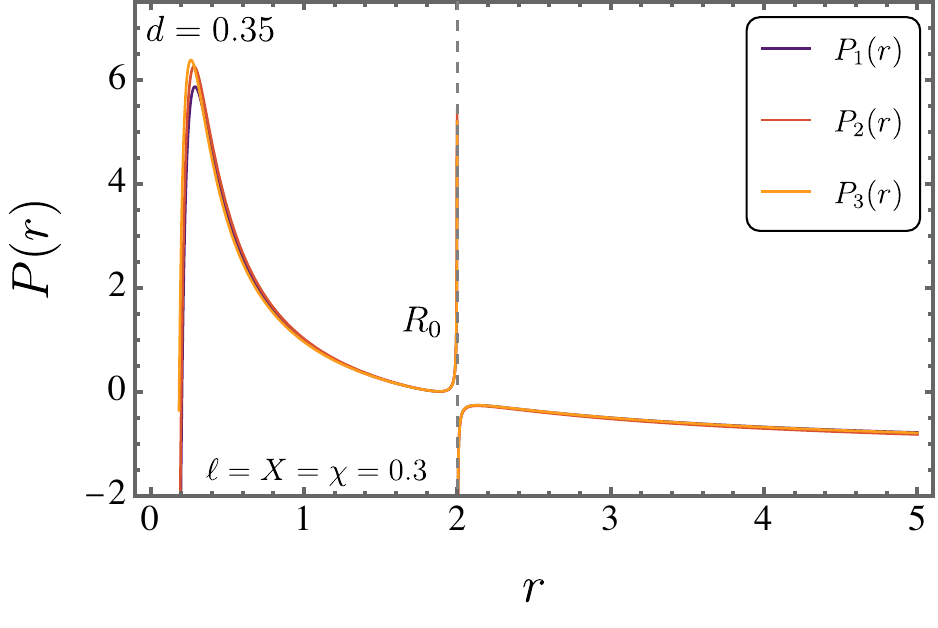}
    \includegraphics[scale=0.44]{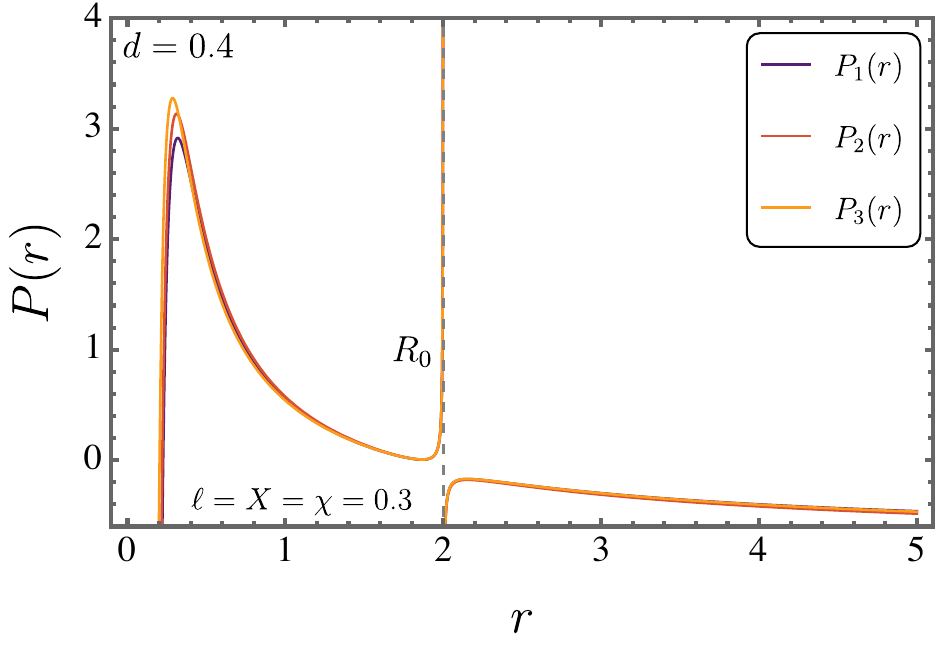}
    \includegraphics[scale=0.44]{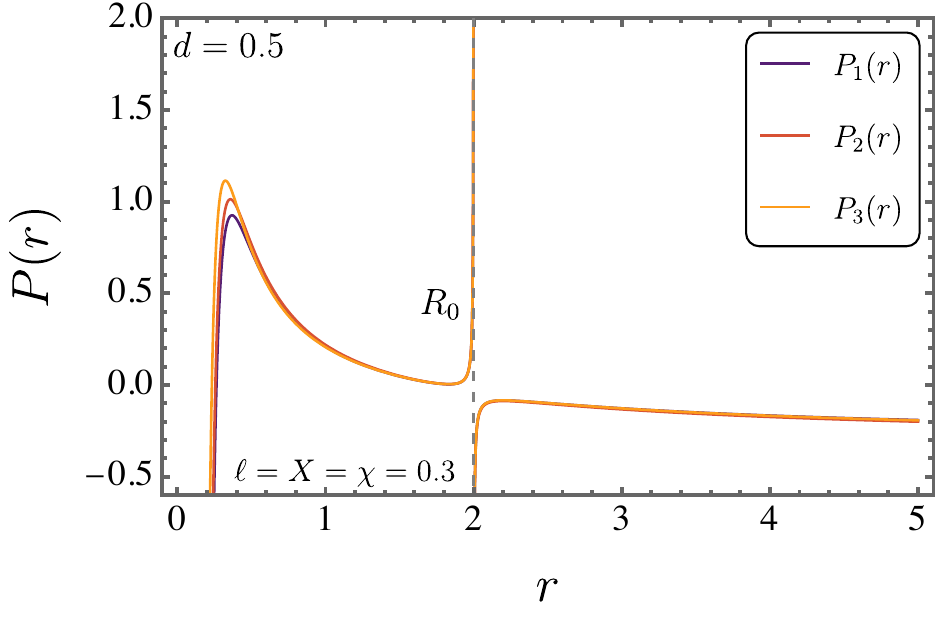}
    \includegraphics[scale=0.44]{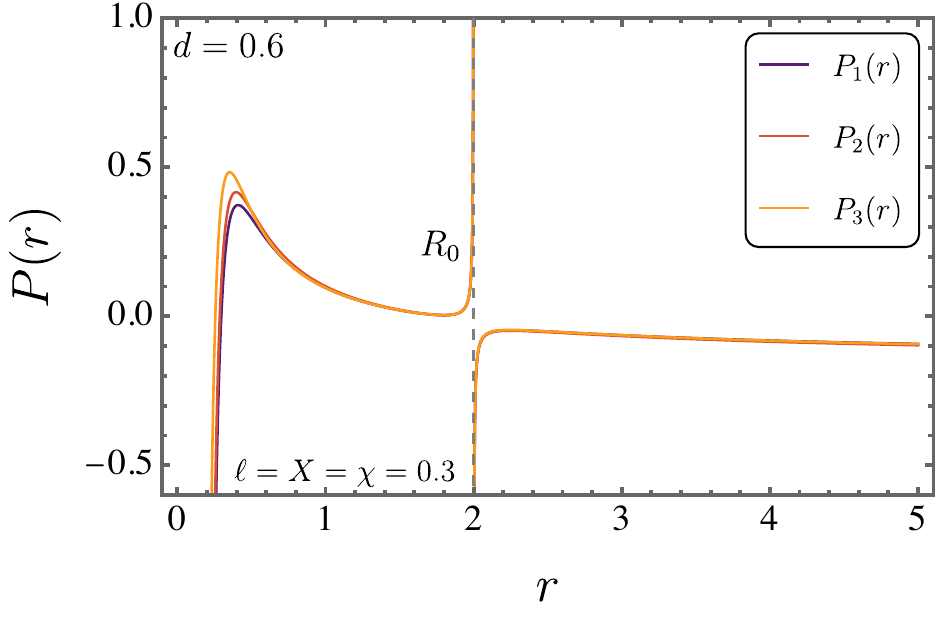}
    \caption{Casimir pressure $P_i(r)$ as a function of the radial coordinate $r$ for the three bumblebee geometries, with fixed plate separation $d$. The parameters are set to $M=1$, $\xi=1/4$, and $\ell=X=\chi=0.3$. As $r\to R_{0}=2M$, all configurations exhibit a divergence behavior.}
    \label{fipressure1}
\end{figure}

\begin{figure}
    \centering
    \includegraphics[scale=0.44]{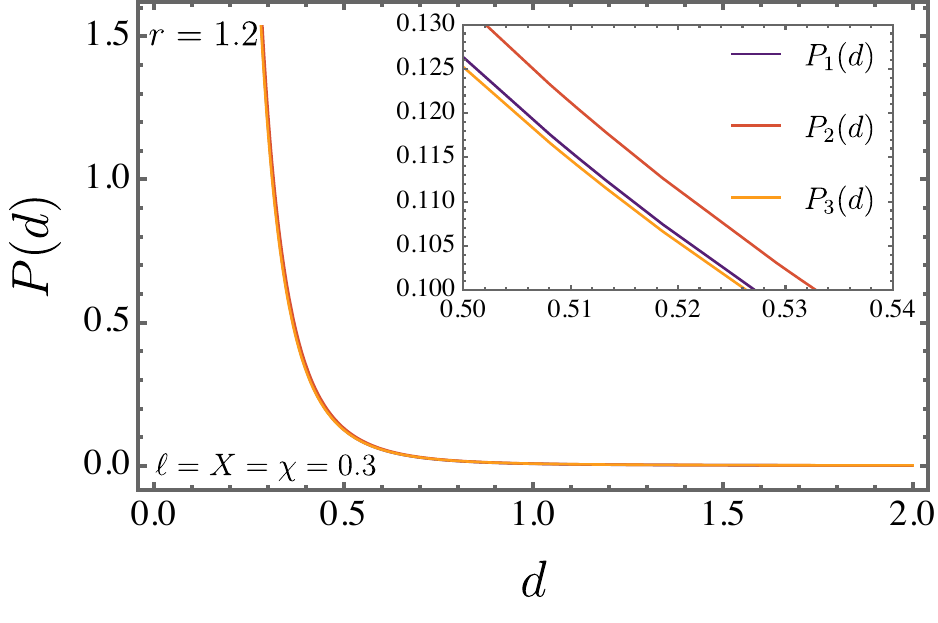}
    \includegraphics[scale=0.44]{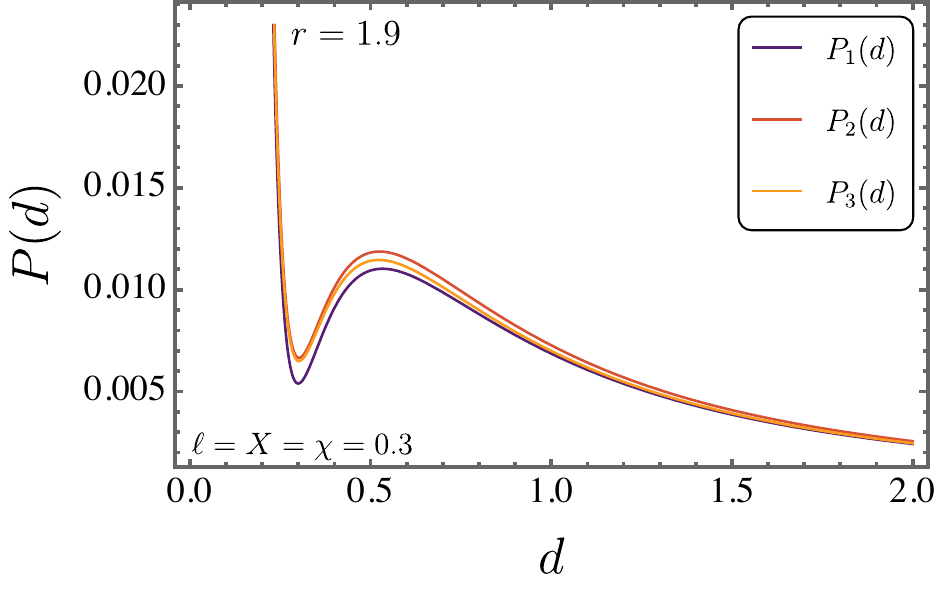}
    \includegraphics[scale=0.44]{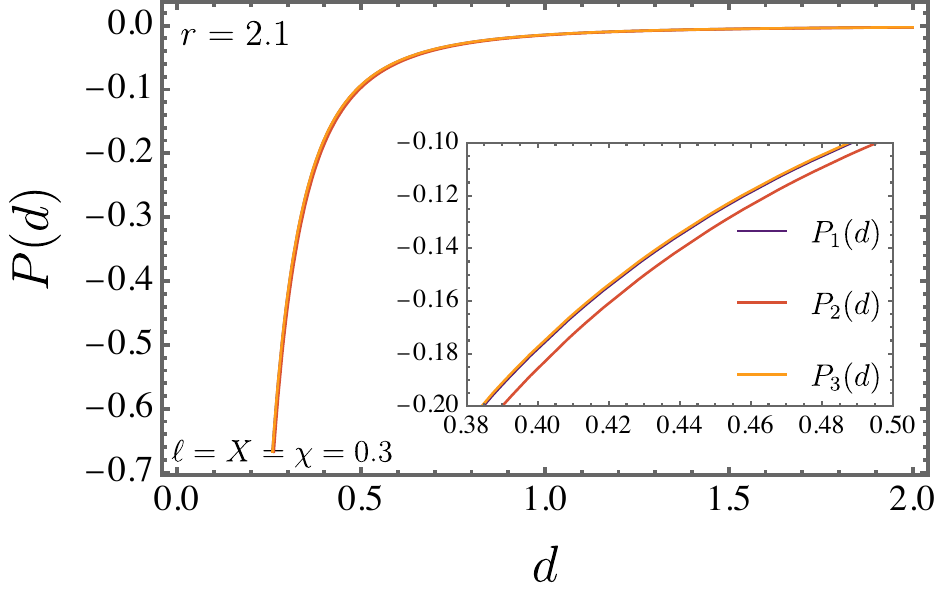}
    \includegraphics[scale=0.44]{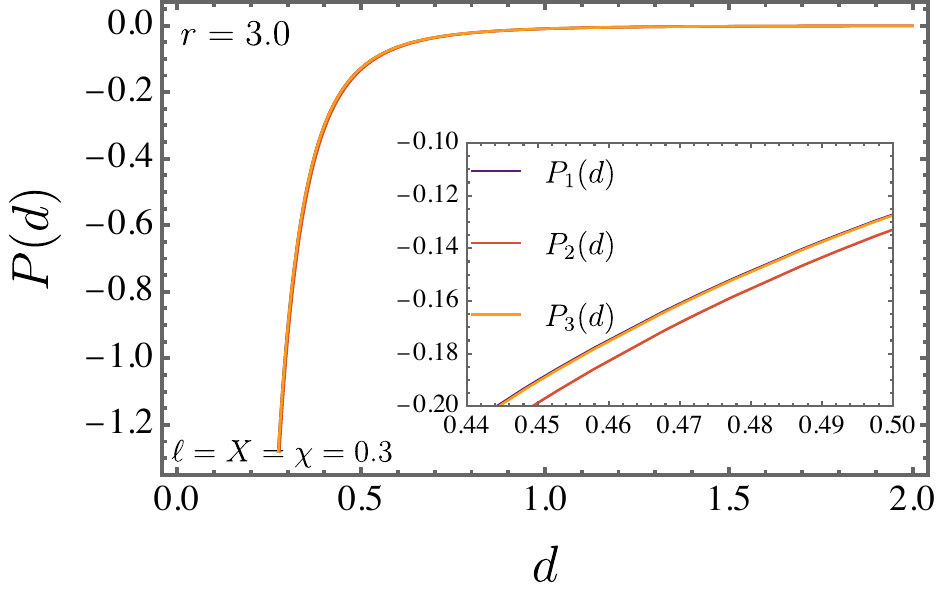}
    \caption{Casimir pressure $P_i(d)$ as a function of the plate separation $d$ for fixed radial distance $r$, considering $M=1$, $\xi=1/4$, and $\ell=X=\chi=0.3$.}
    \label{fipressure2}
\end{figure}

For the pressure, the analysis follows the same structure as in the energy case and is illustrated in Figs.~\ref{fipressure1} and \ref{fipressure2}. The first figure presents the behavior for fixed plate separation $d$, whereas the second displays the pressure as a function of $d$ for fixed radial distance $r$, considering two representative values of the bumblebee parameters $\ell$, $X$, and $\chi$.

When $r<R_{0}$ and $d$ is fixed, the hierarchy
$P_{3}(r) > P_{2}(r) > P_{1}(r)$
is observed. This ordering mirrors the pattern found for the Casimir energy and indicates that the combined vacuum configuration with $\ell=X=\chi=0.3$ enhances the magnitude of the pressure relative to the other geometries. A relevant feature in this regime is the divergence at $r=R_{0}$, which reflects the singular structure inherited from the underlying spacetime background.

In contrast, when $r$ is fixed and the pressure is analyzed as a function of $d$, the behavior changes qualitatively across the surface $r=R_{0}$. For $r<R_{0}$ (for instance, $r=1.2$), the ordering becomes
$P_{2}(d) > P_{1}(d) > P_{3}(d)$,
whereas for $r>R_{0}$ the hierarchy reverses to $P_{1}(d) > P_{3}(d) > P_{2}(d)$. In other words, this inversion shows that non--metricity affects the pressure differently depending on the radial domain. Outside the characteristic radius $R_{0}$, the modifications systematically reduce the magnitude of the pressure, while inside this region the relative dominance among configurations is altered.

%%%%%%%%%%%%%%%%%%%%%%%%%%%%%%%%%%%%%%%%%%%%%%%%%%%%%%%%%%%%%%%%%%%%%%%%%%%%%%%%%%%%%%%%%%%%%%%%%%%%%%%%%%%%%%%%%%%%%%%%%%%%%%%%%%%%%%%%%%%%%%%%%%%%%%%%%%%%%%%%%%%%%%%%%%%%%%%%%%%%%%%%%%%%%%%%%%%%%%%%%%%%%%%%%%%%%%%%%%%%%%%%%%%%%%%%%%%%%%%%%%%%%%%%%%%%%%%%%%%%%%%%%%%%%%%%%%%%%%%%%%%%%%%%%%%%%%%%%%%%%%%%%%%%%%%%%%%%%%%%%%%%%%%%%%%%%%%%%%%%%%%%%%%%%%%%%%%%%%%%%%%%%%%%%%%%%%%%%%%%%%%%%%%%

\section{Conclusion}\label{Sec6}

In this work, we examined the Casimir effect produced by a massive, non--minimally coupled scalar field propagating in three static, spherically symmetric black hole backgrounds arising in bumblebee gravity. Finite–size effects were implemented through the Thermo Field Dynamics formalism by compactifying the radial direction, which allowed us to construct renormalized expressions for the energy–momentum tensor and to derive closed analytic forms for the Casimir energy and pressure in the massless limit.

Although the three geometries share the same asymptotic Schwarzschild structure and horizon radius $R_{0}=2M$, their distinct vacuum realizations of the Lorentz–violating vector field generate quantitatively different vacuum responses. The Casimir energy $E_i(r,d)$ and pressure $P_i(r,d)$ depend explicitly on the radial position of the spherical capacitor, the plate separation $d$, and on the parameters $\ell$, $X$, and $\chi$ that encode the structure of each solution. In the weak–field regime, $r\gg R_{0}$, all configurations recover the standard flat–space scaling proportional to $-1/d^{4}$, confirming the internal consistency of the formalism.

Near the horizon, a universal feature emerges: the Casimir energy vanishes as $r\to R_{0}$, while the radial pressure diverges, with opposite signs when the surface is approached from inside or outside. This behavior reflects the singular character of the horizon for the radial stress component and is inherited from the underlying spacetime geometry.

The most relevant distinctions appear when comparing the hierarchy among the three configurations in different radial domains. Inside the horizon ($r<R_{0}$), the energy ordering satisfies
$E_{2}(d) < E_{1}(d) < E_{3}(d)$,
indicating that the metric–affine solution — where non–metricity modifies the spacetime structure — produces the strongest attractive Casimir interaction in this region. This shows that non–metricity enhances the magnitude of the vacuum energy inside the horizon relative to the purely metric deformation. By contrast, outside the horizon ($r>R_{0}$), the ordering reverses to
$E_{3}(d) > E_{1}(d) > E_{2}(d)$,
revealing that the third configuration, which incorporates simultaneous temporal and radial deformations and represents the most general vacuum structure among the three, maintains the largest deviation from the Schwarzschild behavior in the exterior region.

A similar pattern, with characteristic inversions, is observed for the pressure. For fixed $d$ and $r<R_{0}$, one finds $
P_{3}(r) > P_{2}(r) > P_{1}(r)$,
whereas the hierarchy changes when the pressure is analyzed as a function of $d$ at fixed $r$, and again reverses across the surface $r=R_{0}$. These inversions demonstrate that non–metricity and the more general bumblebee vacuum configurations affect the stress distribution in a domain–dependent manner. In particular, the metric–affine solution plays a dominant role in amplifying the interior vacuum energy, while the third configuration tends to dominate in the exterior region due to its more general modification of both $g_{tt}$ and $g_{rr}$.

For small Lorentz–violating couplings, the three models display similar qualitative profiles, differing mainly in magnitude. As the parameters $\ell$, $X$, and $\chi$ increase, however, the distinction between purely metric and metric–affine realizations becomes more pronounced, especially in the radial dependence of $E_i$ and in the ordering of $P_i$ across the horizon. In general, stronger Lorentz–violating couplings soften the overall magnitude of the Casimir interaction, while preserving the characteristic near–horizon structure.

Therefore, these results show that the Casimir effect is sensitive to both boundary conditions/curvature, and also to the geometric origin of Lorentz symmetry breaking. In particular, non–metricity leaves remarkable signatures on the magnitude and hierarchy of vacuum energy and pressure in strong–field regimes. Extending this analysis to other symmetry–breaking scenarios — such as Kalb–Ramond gravity or non–commutative extensions of the bumblebee framework — may provide further examples in which distinct geometric sectors produce measurable modifications in finite--size quantum effects in curved spacetime.

%%%%%%%%%%%%%%%%%%%%%%%%%%%%%%%%%%%%%%%%%%%%%%%%%%%%%%%%%%%%%%%%%%%%%%%%%%%%%%%%%%%%%%%%%%%%%%%%%%%%%%%%%%%%%%%%%%%%%%%%%%%%%%%%%%%%%%%%%%%%%%%%%%%%%%%%%%%%%%%%%%%%%%%%%%%%%%%%%%%%%%%%%%%%%%%%%%%%%%%%%%%%%%%%%%%%%%%%%%%%%%%%%%%%%%%%%%%%%%%%%%%%%%%%%%%%%%%%%%%%%%%%%%%%%%%%%%%%%%%%%%%%%%%%%%%%%%%%%%%%%%%%%%%%%%%%%%%%%%%%%%%%%%%%%%%%%%%%%%%%%%%%%%%%%%%%%%%%%%%%%%%%%%%%%%%%%%%%%%%%%%%%%%%%

\section{Acknowledgments}

\hspace{0.5cm}
A. A. Araújo Filho is supported by Conselho Nacional de Desenvolvimento Cient\'{\i}fico e Tecnol\'{o}gico (CNPq) and Fundação de Apoio à Pesquisa do Estado da Paraíba (FAPESQ), project numbers 150223/2025-0 and 1951/2025. This work by A. F. Santos is partially supported by National Council for Scientific and Technological Development - CNPq project No. 312406/2023-1. D. S. Cabral acknowledges CAPES for all the financial support provided.

%%%%%%%%%%%%%%%%%%%%%%%%%%%%%%%%%%%%%%%%%%%%%%%%%%%%%%%%%%%%%%%%%%%%%%%%%%%%%%%%%%%%%%%%%%
\section{Data Availability Statement}

Data Availability Statement: No Data associated in the manuscript

%%%%%%%%%%%%%%%%%%%%%%%%%%%%%%%%%%%%%%%%%%%%%%%%%%%%%%%%%%%%%%%%%%%%%%%%%%%%%%%%%%%%%%%%%%

\bibliographystyle{ieeetr}
\bibliography{main}

\end{document}